\documentstyle[galley,prl,aps,psfig]{revtex}

\def\sss{\scriptscriptstyle}
\def\lsim{\hbox{\lower .8ex\hbox{$\, \buildrel < \over \sim\,$}}}
\def\gsim{\hbox{\lower .8ex\hbox{$\, \buildrel > \over \sim\,$}}}

\newlength{\figwidth}
\title{Solitonic transmission of  Bose-Einstein matter waves}
\author{P. Leboeuf, N. Pavloff and S. Sinha\cite{address}}
\address{Laboratoire de Physique Th\'eorique et Mod\`eles Statistiques,
B\^at. 100, \\ Universit\'e Paris-Sud, F-91405 Orsay Cedex, France}

\begin{document}

\maketitle {\begin{abstract} We consider a continuous atom laser propagating
through a wave guide with a constriction. Two different types of transmitted
stationary flow are possible. The first one coincides, at low incident
current, with the non-interacting flow. As the incident flux increases, the
repulsive interactions decrease the corresponding transmission coefficient.
The second type of flow only occurs for sufficiently large incident currents
and has a solitonic structure. Remarkably, for any chemical potential there
always exists a value of the incident flux at which the solitonic flow is
perfectly transmitted.
\end{abstract}}
{}
\narrowtext

\pagebreak
\tightenlines

The transport properties of matter confined to small structures display
distinct quantum effects qualitatively different from those observed at
macroscopic scales. These are grounded on global phase coherence throughout
the sample and can be, in many cases, understood within single particle
pictures, without referring to any specific details of the system. As a
result, they arise in many different fields (electronic systems, atomic
physics, electromagnetism, acoustics) \cite{She95,Imr97,Rai97}. Some examples
are (weak and strong) localization, Bloch oscillations and conductance
quantization.

	Recent experimental developments in the physics of Bose-Einstein
condensation (BEC) of dilute vapor (in particular the microchip guiding
technique) open up the prospect of studying coherent transport phenomena using
guided atom lasers \cite{Lea02}. Besides, because of the extraordinary control
over these systems, they offer a unique opportunity to go beyond the single
particle behavior, and to study specific effects induced by interaction. In
the present article we focus on a simple situation, where a BEC matter wave
propagates through a guide with a constriction ~\cite{Thy99}. By an adiabatic
approximation, the three--dimensional flow is reduced to one dimension, where
the atoms now feel, due to the constriction, a longitudinal step--like
potential of height $V_0$. In the absence of interaction, the transmission $T$
does not depend on the incident current but only on the beam's energy; $T$ is
always lower than unity and tends to this limit when the energy of the beam is
large compared to $V_0$. In the following we consider atoms with a repulsive
effective interaction characterized by a scattering length $a_{sc}>0$. The
most salient features of the flow are all at variance with respect to the
non-interacting case: (i) the transmission coefficient depends on the current,
(ii) at given chemical potential, there exists a maximum transmitted current
above which no stationary flow exists, (iii) at a given current, several 
distinct stationary
solutions with different $T$ are possible and (iv) for any chemical potential
larger than $V_0$, there is a particular value of the incident current
which induces total transmission.

	Consider a continuous atom laser incident on a constriction of a waveguide.
Within the adiabatic approximation \cite{Gla88} the transverse motion is
restricted to the lowest transverse eigen-state. The constriction affects the
longitudinal motion via an effective step-like potential whose magnitude is
fixed by the ground state energy of the transverse Hamiltonian. For a BEC
system, the adiabatic approximation implies that the condensate wave function
can be cast in the form (see Ref.~\cite{Jac98}),
\begin{equation}\label{adia}
\Psi(\vec{r},t)=\psi(x,t)\phi(\vec{r}_\perp;n;x) \; ,
\end{equation}
\noindent where $\psi(x,t)$ describes the motion along the axis
of the laser (the beam is flowing along the positive $x$ direction). $\phi$ is
the equilibrium wave function (normalized to unity) in the transverse
$(\vec{r}_\perp)$ direction. It depends parametrically on the longitudinal
density $n(x,t)=\int d^2r_\perp |\Psi|^2=|\psi(x,t)|^2$. The beam is confined
in the transverse direction by a trapping potential
$V_\perp(\vec{r}_\perp;x)$, which is $x$-dependent in the region of the
constriction. Then, the longitudinal wave equation reads \cite{Jac98,Leb01}
(in units where $\hbar=m=1$)
\begin{equation}\label{e1}
-\frac{1}{2}\partial_{xx}\psi + 
\Big\{V_\parallel(x) + \epsilon[n(x,t);x]\Big\}\,\psi =
 i\,\partial_t\psi \; .
\end{equation} 
  In (\ref{e1}), $V_\parallel(x)$ represents an effective longitudinal
potential due to the constriction. If, to be specific, we consider a
transverse harmonic confinement with pulsation $\omega_\perp(x)$, then
$V_\parallel(x)=\omega_\perp(x)-\omega_\perp(-\infty)$ (energy is measured
with respect of the ground state energy of the non interacting transverse
Hamiltonian far before the constriction). $\epsilon(n;x)$ is a nonlinear term
describing the mean field interaction averaged over a transverse slice of the
beam. One has $\epsilon(n;x) = 2\omega_\perp(x) \, na_{sc}$ in the low density
regime ($n a_{sc}\ll 1$), and $\epsilon(n;x) = 2\omega_\perp(x)\sqrt{na_{sc}}$
in the high density regime ($n a_{sc}\gg 1$) \cite{Jac98,Leb01}.

   Our purpose is to determine the transmission of steady state
solutions of (\ref{e1}) where $\psi(x,t)= \exp\{-i\mu t\} A(x)\exp\{i S(x)\}$,
with $A$ and $S$ real functions. The density is $n=A^2$ and the local
velocity is $v=dS/dx$. From Eq.~(\ref{e1}) one
obtains (i) flux conservation: $n(x)v(x)$ is a constant that we denote
$J_\infty$, and (ii) a Schr\"odinger-like equation for the amplitude:
\begin{equation}\label{e2}
-\frac{1}{2}\, \frac{d^2A}{dx^2}
+ \left\{ V_\parallel(x) + \epsilon[n(x);x] + \frac{J_\infty^{2}}{2\, n^2(x)}
 \right\} A = \mu \, A \, .
\end{equation}
To define the scattering problem one needs to
study the asymptotic behavior of the flow far from the constriction. Far
upstream $V_\parallel(x\to -\infty)=0$ and
the non-linear term in (\ref{e1}) looses its explicit $x$ dependence, taking
the simpler form $\epsilon[n(x,t)]$.
Thus, in this region, (\ref{e2}) admits a first integral of the form
\cite{Leb01}~: 
\begin{eqnarray}\label{e3}
\frac{1}{2} \left(\frac{dA}{dx}\right)^2 
+ W[n(x)] & = & E_{cl}\; \\ 
\nonumber \mbox{with}\;
W(n) & = & -\varepsilon(n) + \mu \, n +\frac{J_\infty^2}{2\, n}\; ,
\end{eqnarray}
\noindent where $\varepsilon(n)=\int_0^n \epsilon(\rho)d\rho$ and $E_{cl}$ is
an integration constant. Eq.~(\ref{e3}) has a simple interpretation in terms
of classical dynamics. It expresses the energy conservation of a
fictitious classical particle with
``position'' $A$ and ``time'' $x$, moving in a potential $W$; $E_{cl}$
being the total energy of this particle. Eq.~(\ref{e3}) is thus integrable by
quadrature, and the density profile can be deduced from the
plot of Fig.~1. Small values of $E_{cl}-W(n_1)$ correspond to small
density oscillations, whereas the highest acceptable value is
$E_{cl}=W(n_2)$, corresponding to a gray soliton.

\begin{figure}\label{fig1}
\centerline{\psfig{figure=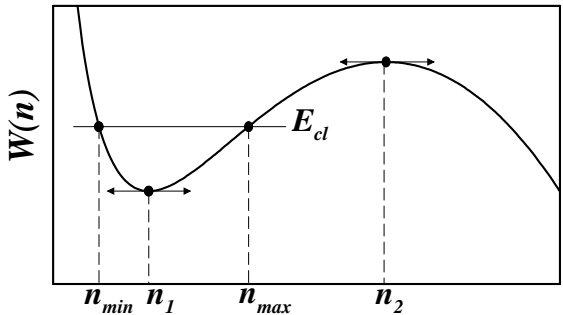,width=\figwidth}}
\caption{$W$ as a function of $n$. $n_1$ and $n_2$ are the zeros of $dW/dn$.
At a given $E_{cl}$, the up-stream density (or equivalently the ``position''
of the fictitious particle) oscillates between $n_{min}$ and $n_{max}$ defined
by $W(n_{min})=W(n_{max})=E_{cl}$.}
\end{figure}

	In the far down-stream region, $V_\parallel(x\to +\infty)$ also takes a
constant value $V_0=\omega_\perp(+\infty) - \omega_\perp(-\infty)>0$. Hence,
(\ref{e2}) admits, in this region, a first integral analogous of (\ref{e3})
where, due to the change in $\omega_\perp$, $\epsilon(n)$ (resp.
$\varepsilon(n)$) takes a different form which we denote $\epsilon_0(n)$
(resp. $\varepsilon_0(n)$). The new form of $W(n)$ is denoted $W_0(n)$ and the
new constant of integration is $E_{cl}^0$:
\begin{eqnarray}\label{e3bis}
\frac{1}{2} \left(\frac{dA}{dx}\right)^2 
+ W_0[n(x)] & = & E^0_{cl}\; \\ 
\nonumber \mbox{with}\;
W_0(n) & = & -\varepsilon_0(n) + (\mu-V_0) \, n +\frac{J_\infty^2}{2\, n}\; .
\end{eqnarray}
 It follows from general arguments on the dispersion of elementary excitations
of Eq.~(\ref{e1}) that the physically acceptable boundary conditions of
Eq.~(\ref{e2}) correspond to a constant far down-stream density (see
\cite{Leb01}). The asymptotic $x\to+\infty$ density should thus be equal
either to $n_{1,0}$ (we denote this as ``case A''), or to $n_{2,0}$ (case B);
$n_{1,0}$ and $n_{2,0}$ -- being the analogous of $n_{1}$ and $n_{2}$ of
Fig.~1 -- are extrema of $W_0(n)$. They are solutions of
\begin{equation}\label{e4}
\mu=\epsilon_0(n)+V_0+\frac{J_\infty^2}{2\,n^2} \; .
\end{equation}
In the non-interacting case the term $\varepsilon_0(n)$ is absent from
$W_0(n)$ which has only one minimum ($n_{10}$). Case A is therefore the only
possible solution in non-interacting systems. Case B describes new
non-perturbative effects related to interaction. It corresponds to an
asymptotic down-stream density which is part of a gray soliton. 

	There exists a maximum value $J_\infty^{max}$ of $J_\infty$ above which
Eq.~(\ref{e4}) admits no solution: as $J_\infty$ is increased (keeping $\mu$
and $V_0$ fixed), the two extrema of $W_0 (n)$ move toward each other, until
they coalesce and disappear. This marks the onset of a time--dependent flow.
If, to be specific, we consider the case $\epsilon_0(n)=g_0\,n^{\nu_0}$, then
\begin{equation}\label{e4bis}
J_\infty^{max} = \left[ \frac{2}{\nu_0+2} (\mu-V_0)
\right]^{\frac{1}{\nu_0}+\frac{1}{2}} 
\sqrt{\nu_0}\; g_0^{-1/\nu_0} \; .
\end{equation}
	The scattering process is now well defined. It corresponds to the matching
between two asymptotic densities described by the classical motion of a
particle of energy $E_{cl}$ in a potential $W(n)$ at $x\to-\infty$, and of
energy $E_{cl}^0$ in a potential $W_0(n)$ at $x\to+\infty$ (with $E_{cl}^0$
either equal to $W_0(n_{1,0})$ or to $W_0(n_{2,0})$). Eq.~(\ref{e2}) being
non-linear, an important question is how to properly define a transmission and
a reflection coefficient; i.e., is it possible to disentangle an incident and
a reflected wave in the upstream flow~? We follow here an approach closely
related to usual experimental set-ups, and choose to work with an incident and
a reflected beam which can be approximated by plane waves. This corresponds to
a regime where Eqs.~(\ref{e2},\ref{e3}) can be linearized in the far upstream
region. In this regime, for $x\to-\infty$, we write $n(x)=n_1+\delta n(x)$ and
expand $nW(n)$ to second order in $\delta n$. Then Eq.~(\ref{e3}) leads to
\begin{equation}\label{e5}
\left(\frac{d\, \delta n}{dx}\right)^2 + 
\kappa_1^2 \delta n^2=8 (n_1+\delta n)[E_{cl}-W(n_1)]\; ,
\end{equation}
where $\kappa_1^2=4(v_1^2-c_1^2)$, $v_1=J_\infty/n_1$ being the average
velocity of the
up-stream beam and $c_1= [ n_1 (d\epsilon/dn)_{n_1} ]^{1/2}$ the sound
velocity of a beam with constant density $n_1$. The linearization (\ref{e5})
is valid provided $|\delta n(x)/n_1|\ll \kappa_1^2/c_1^2$. In this regime, if
one further imposes $v_1\gg c_1$, the up-stream density oscillations can be
analyzed in term of incident and reflected {\it particles} (and not
quasi-particles). This allows to unambiguously define the incident, reflected,
and transmitted current as $J_i=(n_1+\delta n_1/2)v_1$, $J_r=
\delta n_1 v_1/2$ and $J_t= n_1 v_1 = J_\infty$ (where $\delta n_1 =
4[E_{cl}-W(n_1)] / \kappa_1^2$). Hence, once $E_{cl}$ is known,
the transmission at given incident current $J_i$ is determined through
\begin{equation}\label{e6}
E_{cl}
= W(n_1)+\frac{\kappa_1^2}{4}\delta n_1 
= W(n_1) +\frac{\kappa_1^2}{2 v_1} J_i (1 - T) \; .
\end{equation}
The linearization procedure explained so far is valid in the case of small
upstream interaction (this is the essence of the condition $v_1\gg c_1$).
However, all interaction effects are fully taken into account in the
down-stream region, where they are indeed more important (the constriction
acts as a barrier which lowers the velocity of the down-stream flow and, by
flux conservation, increases its density \cite{encore}). In the following we
solve the exact non-linear equation (\ref{e2}) and use the linearization
procedure only to define the transmission coefficient $T$. Thus,
the results presented below are of very general validity, but their analysis
in term of transmission coefficient is only correct in so far as the
linearization procedure is valid.

	The method is now the following: for a given $\mu$ and $J_i$,
assume a particular value of $T$. This determines $J_\infty=T\,J_i$,
fixes the form of the function $W_0(n)$, the value of $n(+\infty)$ (it is equal
to $n_{1,0}$ in case A and to $n_{2,0}$ in case B) and of
$E_{cl}^0=W_0[n(+\infty)]$. Integrating (\ref{e2}) backwards from $x=+\infty$
to $x=-\infty$ yields $E_{cl}$, which should be compatible with (\ref{e6}). If
not, the value of $T$ has to be modified until self-consistency is achieved.

	To understand the physical picture we consider an abrupt step-like
constriction. In this geometry, numerical integration of (\ref{e2}) can be
bypassed because $E_{cl}$ is simply expressed in terms of $E_{cl}^0$ (see
Eq.~(\ref{e7})). However, the adiabatic approximation (\ref{adia}) is based on
the assumption that the typical longitudinal length scale is much larger than
the transverse one. This is clearly violated by an abrupt constriction.
Nevertheless, in certain parameter ranges, the adiabatic approximation remains
valid. In order to illustrate this point we compute the transmission of {\it
non-interacting atoms} \cite{ref_adia}, for which the exact solution can be
obtained numerically. We thus consider a linear wave moving in a guide with
harmonic confinement whose transverse pulsation changes abruptly (at $x=0$
say) from $\omega_\perp^{<}$ (up-stream) to $\omega_\perp^{>} = \alpha\,
\omega_\perp^{<}$ (down-stream), with $\alpha>1$. The incident atoms occupy
the transverse ground state of the up-stream potential. The numerical solution
of the problem can be worked out by a straightforward 3D generalization of the
procedure devised in Ref.~\cite{sza89} for studying a similar 2D problem. We
denote the transmission $T^{\sss L}$, the superscript recalling that we are in
a linear (i.e., non-interacting) regime. The result for $T^{\sss L}$ is
presented in Fig.~2 for $\alpha=2$ and $\alpha=3$. The vertical bars indicate
the location of the energies of the reflected (thin lines) and transmitted
(thick lines) channels. Conservation of angular momentum along the
longitudinal axis imposes selection rules between channels. These rules
effectively forbid half of the energetically allowed channels, and those
henceforth do not play any role in the transmission.

  Within the adiabatic approximation, the constriction is described by an
abrupt longitudinal step potential of height $V_0=\omega_\perp^{>} -
\omega_\perp^<$. The corresponding transmission is $T^{\sss
L}_{adia}=4[\mu(\mu-V_0)]^{1/2} (\sqrt{\mu}+\sqrt{\mu-V_0})^{-2}$ (represented
by a dashed curve in Fig.~2). The onset of transmission occurs at $\mu= V_0$,
i.e., $\mu/\omega_\perp^< = \alpha-1$. One notices in the figure that when
$\mu$ is increased from this value, the adiabatic approximation is initially
quite accurate. However, at larger values of $\mu$ deviations from
adiabaticity are clearly visible. For $\alpha=2$, a sudden lowering of $T^{\sss
L}$ occurs when a new reflected channel opens at $\mu/\omega_\perp^<=2$ (in
all the following we denote as ``open channels'' those allowed by energy
conservation and symmetry rules). From there on, $T^{\sss L}$ diminishes until
a new transmission channel opens, at $\mu/\omega_\perp^<=5$. At this point, a
sudden increase of $T^{\sss L}$ is observed. At large values of $\mu$,
$T^{\sss L}$ tends to unity, as it should. For $\alpha=3$, the process is
similar, but the breakdown of the adiabatic approximation occurs earlier
because the opening of the initial transmitted channel -- at
$\mu/\omega_\perp^<=2$ -- coincides with that of the first allowed excited
reflected channel. 

  In the following we will concentrate on the region $\mu/\omega_\perp^< \gsim
\alpha-1$ where the adiabatic approximation is well justified and where, as we
shall now see, non-linear effects may induce strong modifications of the
transmission.

\begin{figure}\label{fig2}
\centerline{\psfig{figure=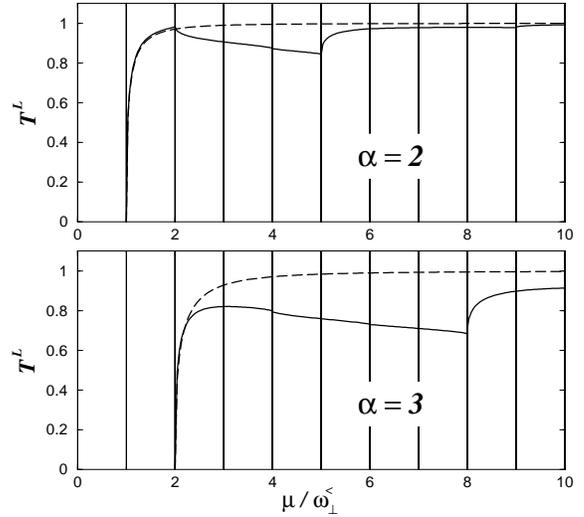,width=\figwidth}}
\caption{Solid curve: transmission $T^{\sss L}$ of a linear wave as a function
of $\mu$ for $\alpha=2$ (top) and $\alpha=3$ (bottom). Dashed curve:
adiabatic approximation. In each plot the vertical thick (thin) lines indicate
the location of the transverse eigen-energies in the down-stream (up-stream)
confining potential.}
\end{figure}

	We now turn to the non-linear problem and consider the case
$V_\parallel(x<0)=0$, $\epsilon(n;x<0)=\epsilon(n)$ and
$V_\parallel(x>0)=V_0$, $\epsilon(n;x>0)=\epsilon_0(n)$. Eq.~(\ref{e2}) admits
the first integral (\ref{e3}) for all $x\le 0$ and (\ref{e3bis}) for all $x\ge
0$. $E_{cl}$ is determined through $E_{cl}^0$ by imposing continuity of $A$
and $A'$ at $x=0$:
\begin{equation}\label{e7}
E_{cl}-W[n(0)]=E_{cl}^0-W_0[n(0)]\; .
\end{equation}
	Let's consider case A first. The asymptotic down-stream density is $n_{1,0}$
and thus one has, for all $x\ge 0$, $n(x)=n_{1,0}$ (the fictitious classical
particle remains at the bottom of the potential well $W_0$). In particular,
$n(0)=n_{1,0}$ and the matching (\ref{e7}) determines $E_{cl}$ uniquely. The
value of $T$ is denoted $T^{\sss A}$ in this case.

	Case B is more interesting because the structure of the down-stream solution
is richer: $n(x\ge 0)$ being part of the profile of a gray soliton, $n(0)$ can
be varied continuously provided the matching (\ref{e7}) is fulfilled at a
value acceptable for Eq.~(\ref{e3}). Effectively, the only restriction imposed
is that $n_{min}<n(0)<n_{max}$ ($n_{min}$ and $n_{max}$ are defined in
Fig.~1). As a result, for fixed $\mu$ and $J_i$, $E_{cl}$ is not
uniquely determined by $n(+\infty)$, and the transmission $T^{\sss B}$ varies
between 0 and a value that we denote as $T^{\sss B}_{max}$.

	To be specific, we consider a continuous beam of $^{23}$Na atoms propagating
through a guide with a transverse confinement $\omega_\perp^<=2\pi\times 2$
kHz (in the region $x<0$), to which we impose a narrowing
$\omega_\perp^>=2\pi\times 6$ kHz (in the region $x>0$). This represents a
barrier of height $V_0=192$ nK. In the non-interacting case the transmission
$T^{\sss L}$ as a function of $\mu$ is plotted in the bottom part of Fig.~2
($\alpha = 3$). We take $\mu=210$ nK (this corresponds to the kinetic energy
of atoms having a velocity of 1.2 cm/s). This value of $\mu$ corresponds, in
the non-interacting case, to a regime where the adiabatic approximation holds
and yields a transmission $T^{\sss L}\simeq T^{\sss L}_{adia}\simeq 0.70$. The
repulsive interaction between atoms introduces in Eq.~(\ref{e2}) a non-linear
term which, in the region $x<0$, reads $\epsilon(n)=gn$ with $g= 2\,
a_{sc}\,\omega_\perp^< \,\simeq 530$ nK.nm. In the region $x>0$, the transverse
frequency of the guide is multiplied by 3, and thus $\epsilon_0(n)=g_0n$ with
$g_0=3\times g$. An important parameter of the system is the maximum
transmitted current $J_\infty^{max}$ above which no stationary flow can exist
in the down-stream part of the guide. From Eq.~(\ref{e4bis}) (with $\nu_0=1$)
one obtains $J_\infty^{max} \simeq 1.6\times 10^4$ atom/s.

	Fig.~3 summarizes the results obtained. The linearization condition $v_1\gg
c_1$ is extremely well satisfied in the whole range of incident currents
considered (the less favorable case occurs at large $J_i$, where $v_1\simeq
25\,c_1$). The horizontal dashed line is the value of the transmission
coefficient of non-interacting atoms~:  $T^{\sss L}\simeq 0.7$. It is current
independent. When the current is increased from zero, $T^{\sss A}$ decreases
from this value down to $T^{\sss A}\simeq 0.5$. At this point (located with a
black spot on the figure), $J_\infty(=T^{\sss A}J_i)$ is equal to
$J_\infty^{max}$, and a stationary flow of type A is no longer permitted
(actually it bifurcates to a type B solution). The prominent feature of the
behavior of $T^{\sss A}$ as a function of $J_i$ is its decrease compared to
the non-interacting value $T^{\sss L}$. The physical reason behind this
phenomenon is simple: the available kinetic energy necessary to step over the
barrier is reduced when the interaction energy increases, i.e., when the
incident current increases. This picture is supported by a perturbative
treatment which accurately describes the flow at low incident current ($J_i\ll
J_\infty^{max}$) and confirms that the decrease of $T^{\sss A}$ corresponds to
an increased fraction of the interaction energy in the chemical potential.

\begin{figure}\label{fig3}
\centerline{\psfig{figure=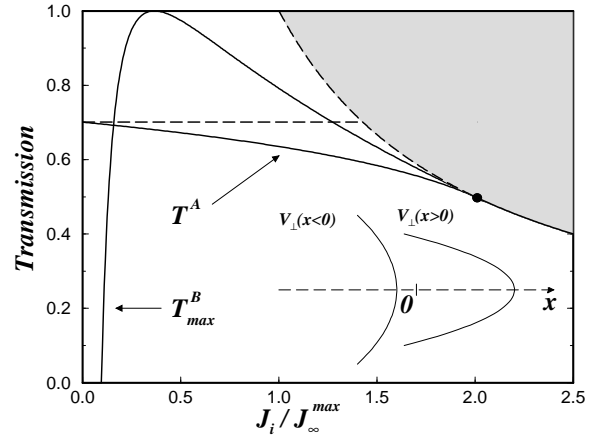,width=\figwidth}}
\caption{$T^{\sss A}$ and $T^{\sss B}_{max}$ as a function of $J_i
/J_\infty^{max}$ (at fixed $\mu=210$ nK). The horizontal dashed line is the
transmission $T^{\sss L}$ of non-interacting atoms. The gray zone above the
dashed hyperbola (of equation $T=J_\infty^{max}/J_i$) is a region where no
stationary flow exists. Inset: Schematic of the constriction's geometry.}
\end{figure}

	Case B being mediated via interaction, does not exist for low current. It
exists only above a critical current $(2g_0)^{-1}(\mu-V_0)^2
(8\mu)^{-1/2} \simeq 1.5\times 10^3$ atom/s $\simeq 0.1 \, J_\infty^{max}$.
From this point, $T^{\sss B}_{max}$ increases rapidly up to 1 (reached at
$J_i\simeq 0.45 \, J_\infty^{max}$ in the case of Fig.~3), and then decreases
down to a point where one can show that it exactly meets the end point of
$T^{\sss A}$. From there on, the value of $T^{\sss B}$ is limited by the
condition that the flow should be stationary, and one has $T^{\sss B}_{max} =
J_\infty^{max}/J_i$, which coincides with the dashed hyperbola in Fig.~3.
We emphasize that
stationary solutions of type $B$ with arbitrary transmission $0 \le T^{\sss B}
\le T^{\sss B}_{max}$ exists for any current above the critical one.

The nonlinear transport induced by the re\-pul\-si\-ve two-body interaction
has therefore a non-trivial consequence: new solutions -- of solitonic
character -- emerge; they allow for an increased transmission. This contrasts
with the behavior of case A where the transmission is lowered by the
interaction. One can show that, for any value of $\mu>V_0$, there always
exists a value of $J_i$ such that complete transmission exists in case B. The
profile for $T^{\sss B}_{max}=1$ consists of a constant up-stream density
$n(x\le 0)=n_1$ connected at $x=0$ to half a soliton. Fig.~4 displays the
density profiles of two stationary flows (case $A$ and case $B$ at $T^{\sss
B}=T^{\sss B}_{max}=1$) at an incident current $J_i=0.36\,J_\infty^{max}$ (see
Fig.~3). Note the significant difference in the densities of the up-stream
and down-stream profiles, a purely nonlinear effect mediated by a solitonic
profile. We have performed numerical computations that show that the same type
of solution also exists for smooth constrictions and that they are dynamically
stable, as confirmed by a Bogoliubov analysis. The interactions can thus have
two different and, in some sense, opposite consequences on the transport
properties of a condensate flow. They diminish the transmission in some
instances (case A), but also allow for new stationary flows that can be
perfectly transmitted (case B) \cite{Kag03}.

\begin{figure}\label{fig4}
\centerline{\psfig{figure=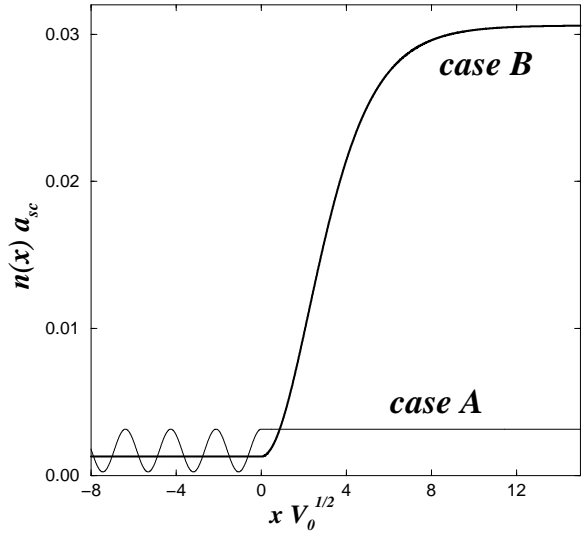,width=\figwidth}}
\caption{Density profiles (in dimensionless units) for the constriction
defined in the text, at incident current $J_i=0.36\,J_\infty^{max}$ (flowing
from left to right) and chemical potential $\mu=210$ nK. Thick solid line:
solitonic flow with $T^{\sss B}=T^{\sss
B}_{max}=1$; thin solid line: type-A flow with $T^{\sss A}=0.7$.}
\end{figure}

   We have restricted our analysis to stationary configurations, but dynamical
effects are certainly of interest. Amongst these, one could address the
question of the transient that exists before stationarity is reached, or
the nature of the flow in parameter regions where stationarity is not
possible. An other open problem, that clearly deserves investigation, is the
dynamical selection of the different types of stationary profiles discussed
above: given an initial low density flow (which, from Fig.~3, is of type A),
which branch (A or B) will be followed when the incident flux increases ?

    There are different ways to experimentally realize the effect discussed in
the present work, namely enhanced solitonic transmission of matter waves. We
have studied one possible implementation, where the step--like potential in the
longitudinal motion of the condensate is produced by a constriction of the
guide. An other possibility is to apply a blue-detuned laser beam on the
region $x > 0$ of a condensate propagating along a guide of constant diameter.
In this case, the characteristics of the flow and the barrier should be easily
controlled by modifying the laser's frequency, intensity and waist, thus
allowing for a neater experimental observation of the above predicted
transport phenomena \cite{enplus}.

	We acknowledge stimulating discussions with D. Gu\'ery-Odelin. L.P.T.M.S. is
Unit\'e Mixte de Recherche de l'Universit\'e Paris XI et du CNRS, UMR 8626.


\begin{references}
\bibitem{encore} Down-stream interaction effects are further amplified
by the effective nonlinear term $\epsilon(n;x)$ which increases with
transverse confinement.
\bibitem{enplus} Besides, the adiabatic approximation is better justified if
the potential step for the linear motion results from an external source
(in the absence of interaction it is rigorously exact since the problem is
separable).
\end{references}
\end{document}